\documentclass[aps,preprint,groupedaddress,showpacs]{revtex4}
\usepackage[dvips]{graphicx}
\usepackage{dcolumn}
\usepackage{amsmath}
\begin{document}

\title{Soft Interaction Between Dissolved Dendrimers: Theory and Experiment}
	   
\author{C.~N. Likos}
\affiliation{Institut f{\"u}r Theoretische Physik II,
Heinrich-Heine-Universit\"at D{\"u}sseldorf,
Universit\"atsstra{\ss}e 1, D-40225 D\"usseldorf, Germany}
\author{M.~Schmidt}
\affiliation{Institut f{\"u}r Theoretische Physik II,
Heinrich-Heine-Universit\"at D{\"u}sseldorf,
Universit\"atsstra{\ss}e 1, D-40225 D\"usseldorf, Germany}
\author{H.~L{\"o}wen}
\affiliation{Institut f{\"u}r Theoretische Physik II,
Heinrich-Heine-Universit\"at D{\"u}sseldorf,
Universit\"atsstra{\ss}e 1, D-40225 D\"usseldorf, Germany}
\author{M.~Ballauff}
\affiliation{Polymer-Institut, Universit\"at Karlsruhe,
Kaiserstra{\ss}e 12, D-76128 Karlsruhe, Germany}
\author{D.~P{\"o}tschke}
\affiliation{Polymer-Institut, Universit\"at Karlsruhe,
Kaiserstra{\ss}e 12, D-76128 Karlsruhe, Germany}
\author{P.~Lindner} 
\affiliation{Institut Laue-Langevin, B.P.\ 156X, 38042 Grenoble Cedex, France} 
\date{Submitted to {\sl Macromolecules} on July 24, 2000}

\begin{abstract}
Using small-angle neutron scattering and liquid integral equation theory,
we relate the structure factor of flexible dendrimers of 4th generation
to their average shape. The shape is measured as a radial density 
profile of monomers belonging to a single dendrimer. From that, we
derive an effective interaction of Gaussian form between pairs of
dendrimers and compute the structure factor using the hypernetted chain
approximation. Excellent agreement with the 
corresponding experimental results is
obtained, without the use of adjustable parameters.
The present analysis thus strongly supports the previous 
finding that flexible dendrimers of low generation present 
fluctuating structures akin to star polymers.
\end{abstract}
\maketitle
\section{Introduction}

\begin{figure}
\begin{center}
\includegraphics[width=15.0cm,height=22.0cm]
{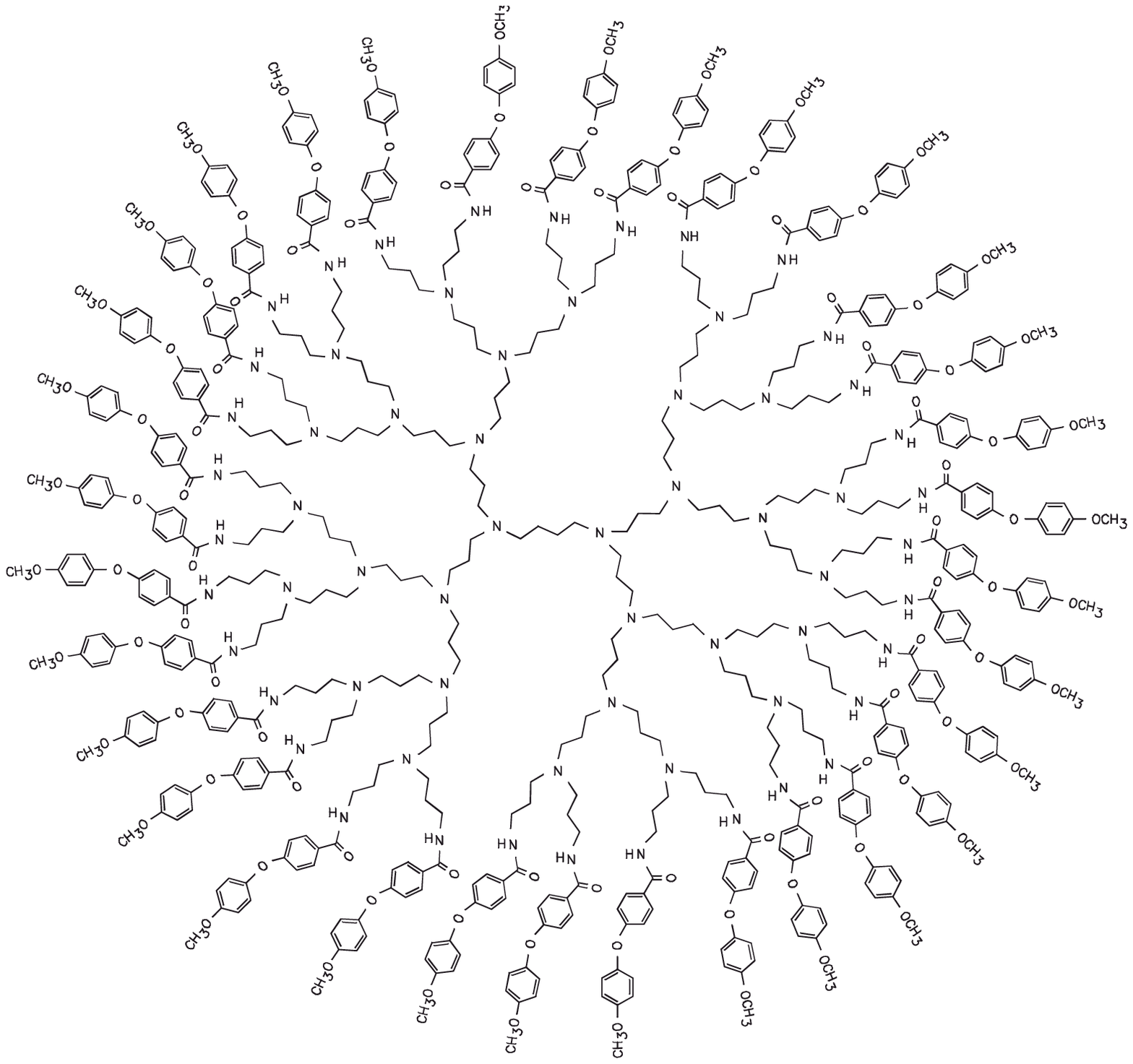}
\caption{Structure of dendrimer G4.}
\label{dendrimer}
\end{center}
\end{figure}

Dendrimers are synthetic branched macromolecules with 
defined structure \cite {Vogtle:99}. Starting from a 
trifunctional flexible monomer (generation 0), 
subsequent shells of trifunctional units 
are  attached in a well-defined manner.
Fig.\ \ref{dendrimer}
displays a dendrimer of 4th generation with attached end groups 
at the ends of the units constituting the last generation. 
Thus, a tree-like structure is generated, which can be viewed as an interesting 
intermediate between colloids and polymers: 
dendrimers of low generations exhibit enough degrees of 
conformational freedom and will thus present fluctuating 
structures in solution. Hence, these structures will greatly 
resemble star polymers or lightly branched polymers. 
If, on the other hand, the number of generations is increased, 
steric interactions between the groups located at the 
periphery of the molecule must result. 
In this case, significant back-folding of these groups 
must occur and these structures will exhibit a more 
homogeneous segmental density. 
Hence, dendrimers of higher generation will resemble 
dense colloidal structures.

The average radial density distribution $T(r)$ 
of dendrimers is still a matter of debate. 
Here, the question arises, whether dissolved dendrimers exhibit 
their maximum segment density in the center or at the periphery of the 
molecule. The first theoretical analysis of the radial density 
distribution by Hervet and de Gennes 
came to the conclusion that dendrimers have a dense shell 
and hence a minimum of $T(r)$ at the center of the molecule \cite{Hervet:83}. 
Subsequent theoretical studies, however, showed that dendrimers exhibit 
their maximum segment density at the center of the 
molecule \cite{Muthu:90, Mansf:93, Murat:96, Rub:96, Muthu:98, Lue:00}. 
Only if electrostatic repulsion operates between the segments of the dendrimer, a dense shell structure is to be expected \cite{Muthu:98}.
 
Small-angle scattering methods, such as small-angle X-ray scattering 
(SAXS) \cite{Feigin:87} and small-angle neutron scattering 
(SANS) \cite{Higgins:94} are suitable to investigate the radial 
structure of dissolved dendrimers. 
Up to now, however, SANS- and SAXS-studies of dendrimers in solution 
did not yet come to a final conclusion regarding the average 
radial density distribution $T(r)$.
A survey of literature may be found in a recent review \cite{B:00}. 
A recent study of a dendrimer of 7th generation concluded 
that the internal structure of the molecule is rather uniform, 
with the end groups being preferably located at the periphery 
of the molecule \cite{Topp:99}. 
This result is in contradiction to the theoretical 
studies discussed 
above \cite{Muthu:90, Mansf:93, Murat:96, Rub:96, Muthu:98, Lue:00}. 
SANS-studies of dendrimers of lower generation, however, 
showed that the molecules under consideration have approximately 
a Gaussian density distribution \cite{ballauff:99, Potschke:99}. 
The chemical structure of the dendrimers investigated in Refs.\ \cite{Topp:99} 
and \cite{ballauff:99, Potschke:99},  
differ with regard to the number of generations and the 
nature of the end groups. 
The results can therefore only be compared with caution. In this respect,
we also mention the simulation results of Ref.\ \cite{Lue:00} where it
was found that the conformation of dendritic molecules drastically depends
on the generation number, with a soft, Gaussian-like profile resulting
for small generations and a ``collapsed core'' with soft tails for larger
generations. 
Nevertheless, it must be concluded that the question of the 
overall structure of dissolved dendrimers has not yet 
found a generally accepted answer 
(see the discussion of this point in Ref.\ \cite{B:00}.)

Small-angle scattering conducted at different particle 
concentrations may be useful for further elucidation of this problem. 
Neglecting incoherent contributions, the measured scattering 
intensity $I(q)$ ($q$: magnitude of scattering vector $\textbf{q}$; 
$q = (4\pi/ \lambda)\sin(\theta /2) $; $\lambda$: wavelength of radiation;
$\theta$: scattering angle) may be rendered as \cite{Feigin:87, Higgins:94}
\begin{equation}
I(q) = \rho_{\rm d}\left(\bar\rho - \rho_{\rm m}\right)^2I_{\rm S}(q)S(q).
\label{iofq}
\end{equation}
In eq.\ (\ref{iofq}) above,
$\rho_{\rm d} = N_{\rm d}/V$ is the number density of $N_{\rm d}$ 
dissolved dendrimers in the sample volume $V$. The difference
$\bar\rho - \rho_{\rm m}$ is the contrast of the scatterer towards the
solvent, whereby $\bar\rho$ is the average scattering length density of
the scatterer and $\rho_{\rm m}$ that of the solvent.
$I_{\rm S}(q)$ is the form factor of the object, a quantity directly
related to the mass distribution of scattering centers within the
macromolecular aggregate, as will explained below.
Finally,
the structure factor $S(q)$ accounts for all interparticle 
correlations arising from the interactions between those. 

The effective interaction between dendrimer centers
is formally defined as follows \cite{lowen:madden:hansen:93}: 
the centers of mass of the dendrimers
are kept fixed at prescribed positions and a canonical trace is carried
out over all monomer degrees of freedom. This procedure yields a
constrained free energy, depending on the instantaneous configuration
of the dendrimers' centers of mass. Then, the effective interaction
energy between the dendrimers is $-k_BT$ times the logarithm of this
partition function, where $k_B$ is Boltzmann's constant and 
$T$ is the absolute temperature.
When averaging over the dendrimers' positions by employing the
so-defined effective Hamiltonian, 
the thermodynamics of the
system remains invariant. In general, this procedure generates two-, three-
and higher-order interactions between the centers of mass of the dendrimers.
However, we will follow usual practice and limit our considerations
to two-body potentials only, that is, we will make the usual
{\it pair-potential approximation}, introducing an effective
pair interaction $V_{\rm eff}(R)$ between the centers of mass of
two dendrimers separated by the distance $R$. We comment on the
accuracy of the pair potential approximation in section III.
Since we are dealing with particles which have 
an overall spherical shape, this effective interaction is spherically
symmetric and depends only on the magnitude of the vector connecting
the two centers.
 
The structure factor $S(q)$ is directly 
related to the pair correlation 
function $g(R)$ and hence to the 
total interaction potential of the solute molecules \cite{Hansen:86}. 
Once $V_{\rm eff}(R)$ is known, the calculation of 
$g(R)$ and $S(q)$ follows from the solution of any of the rich
variety of so-called liquid integral equation theories \cite{Hansen:86}.
For infinite dilution $S(q) = 1$. Hence, from an experimental
point of view, eq.\ (\ref{iofq}) shows that $I_{\rm S}(q)$ may be 
obtained 
from SANS- or SAXS-data which have been suitably 
extrapolated to vanishing concentration \cite{Higgins:94}. 

The structure factor 
$S(q)$ gives direct information on the steepness 
of the repulsion of the dissolved particles: in the case of hard spheres,
a pronounced structuring of the solution will occur, 
which leads to a marked maximum of $S(q)$ \cite{Hansen:86}. 
If the dissolved particles interact via soft repulsion, 
the maximum of $S(q)$ is decreased. 
Recently, the problem of $S(q)$ of star polymers has been 
addressed \cite{Jus:99, anom}. 
Here, it could be shown that star polymers may be looked upon as 
ultrasoft colloidal particles. 
The predictions of theory have met with gratifying success 
when compared to SANS-data measured for concentrated solutions of star 
polymers \cite{Lik:98,Lik:00}. 

Up to now, only a few experimental 
studies have addressed the problem of the structure 
factor of dissolved dendrimers. Ramzi {\it et al.}\ \cite{Ram:98} 
studied concentrated solutions of dendrimers by SANS. 
In the case of uncharged dendrimers, these workers found a marked 
depression of the peak of $S(q)$ despite the strong decrease 
of the structure factor at low $q$. 
They concluded that the absence of a peak of $S(q)$ 
is related to the softness of interaction, which results from the 
high internal flexibility of the dendritic structures. 
If charges are added to the segments of the dissolved dendrimers, 
a strong peak of $S(q)$ is generated which is clearly traced back 
to the screened Coulombic interactions of the 
now charged species (cf.\ also Ref.\ \cite{Mic:98}.) 
A SANS-study by Topp {\it et al.}\ 
\cite{Amis:99}, 
showed that the dendrimers under consideration exhibited 
only a weak maximum of $S(q)$, in particular at lower generation. 
At high concentrations the evaluation of data according 
to eq.\ (1) did not lead to meaningful results, 
which indicates a considerable change of the internal structure 
of the dendrimers in  this regime. 
The dilute regime has been studied in Refs.\ \cite{ballauff:99, Potschke:99}. 
The data of $S(q)$ have been interpreted in terms of a simplified model,
treating the dendrimers as effective hard spheres. 
It is hence evident that the problem of $S(q)$ of dissolved 
dendrimers is not yet understood quantitatively. 

In this work, we wish to apply a simple theory which allows
us to calculate the effective interaction between dendrimer centers
using the measured density profile of an isolated dendrimer as 
a starting point. It turns out that the interaction has a Gaussian form.
Subsequently, we employ it to
calculate theoretically the  
structure factor $S(q)$ of dissolved dendrimers at various 
concentrations using the hypernetted chain (HNC) approximation. 
The consequences of the internal flexibility on the measured 
structure factor $S(q)$ will be discussed explicitly. 
Moreover, a quantitative comparison with the experimental data presented 
in Ref.\ \cite{Potschke:99} will be given, showing excellent agreement.
It will be demonstrated that the radial density distribution 
$T(r)$ derived from scattering experiments may directly 
be used to calculate $S(q)$.

\section{Theory}

The starting point for the theory is the monomer density profile
$\rho(r)$ around the center of mass of an isolated dendrimer. The
experiments of Ref. \cite{ballauff:99, Potschke:99} offer a direct access to 
this quantity, through the so-called shape function $T(r)$. 
The scattering intensity $I_{\rm S}(q)$ from a solution of dendrimers
at infinite dilution has been written as (see eq.\ (18) in 
Ref.\ \cite{ballauff:99}):
\begin{equation}
I_{\rm S}(q) = T^2(q),
\label{iandt}
\end{equation}
meaning that $T^2(q)$ is the form factor of the dendrimer and hence
$T(q)$ the Fourier transform of $T(r)$. Hence, as already pointed
out in Ref.\ \cite{ballauff:99}, $T(r)$ is a ``shape function'' that
describes the way in which $\rho(r)$ varies in space. This shape
function is dimensionless, hence we may write:
\begin{equation}
\rho(r) = \rho_0T(r),
\label{rhor.1}
\end{equation}
where $\rho_0$ is a constant having dimensions of density (length$^{-3}$)
and which will be specified now. Eq.\ (\ref{rhor.1}) above, together
with eq.\ (2) of Ref.\ \cite{ballauff:99} imply:
\begin{equation}
\int d{\bf r} \rho(r) = \rho_0V_p,
\label{int.rho}
\end{equation}
where $V_p$ is the measured partial volume of the solute molecule. 
On the other hand, the integral of $\rho(r)$ has to yield the total
number of monomers $N$ within the dendrimer and this fact together with
eq.\ (\ref{int.rho}) imply:
\begin{equation}
\rho(r) = \frac{N}{V_p}T(r).
\label{rhor.2}
\end{equation}

The shape function $T(r)$ is determined by an inverse Fourier transform
of $T(q)$, the latter being given by eq.\ (20) of Ref.\ \cite{ballauff:99}.
Ignoring the `tail' $(aq^2 + bq)\exp(-dq^2)$ there, we have:
\begin{equation}
T^2(q) = V_p^2\exp(-q^2R_g^2/3).
\label{tofq}
\end{equation}

\noindent
From eq.\ (\ref{tofq}) and after an inverse Fourier transformation, we
readily obtain:
\begin{equation}
T(r) = V_p\left(\frac{3}{2\pi R_g^2}\right)^{3/2}
              \exp\left(-\frac{3r^2}{2R_g^2}\right).
\label{tofr}
\end{equation}
It can be easily seen that $\int d{\bf r} T(r) = V_p$, in agreement
with eq. (2) of Ref.\ \cite{ballauff:99}. 

From eqs.\ (\ref{rhor.2}) and (\ref{tofr}) we obtain the 
monomer density as:
\begin{equation}
\rho(r) = N\left(\frac{3}{2\pi R_g^2}\right)^{3/2}
              \exp\left(-\frac{3r^2}{2R_g^2}\right).
\label{rhor.3}
\end{equation}

We now assume that two such dendrimers are kept with their centers of
mass at a separation ${\bf R}$ apart and wish to calculate the 
ensuing interaction potential. Let us assume that the monomer-monomer
interaction potential is given by some function $v({\bf r}_1 - {\bf r}_2)$
where ${\bf r}_1$ and ${\bf r}_2$ are the positions of the 
two monomers. In the {\it mean-field approximation}, i.e., 
ignoring the correlations and possible deformations of the dendrimers,
the total interaction potential $V_{\rm eff}({\bf R})$ can be approximated by
a double integral over the unperturbed density profiles times the
monomer-monomer interaction kernel \cite{graf:lowen:98}. 
This approximation should hold
when the dendrimers are not too deeply interpenetrating and reads as:
\begin{equation}
V_{\rm eff}({\bf R}) = \int\int d{\bf r}_1 d{\bf r}_2 \rho({\bf r}_1) 
                      \rho({\bf r}_2 - {\bf R}) v({\bf r}_1 - {\bf r}_2).
\label{dconv}
\end{equation}
We now make the simplest possible assumption and model the monomer-monomer
interaction by a delta function (contact repulsion):
\begin{equation}
v({\bf r}_1 - {\bf r}_2) = v_0 k_BT \delta({\bf r}_1 - {\bf r}_2),
\label{delta}
\end{equation}
introducing the excluded volume parameter $v_0$. Eqs.\ (\ref{dconv})
and (\ref{delta}) then yield:
\begin{equation}
V_{\rm eff}({\bf R}) =  
v_0 k_BT \int d{\bf r} \rho({\bf r}) \rho({\bf r} - {\bf R}).
\label{conv}
\end{equation}
In this approximation, the interaction is proportional to the convolution
of the monomer density of a single dendrimer with itself. But each 
density profile has the form $N(\alpha/\pi)^{3/2}\exp(-\alpha r^2)$,
i.e., $N$ times a normalized Gaussian with width parameter 
$\alpha$.
It is known that the convolution
of a normalized Gaussian having width $\alpha$ with itself is again 
a normalized Gaussian, but with width $\alpha/2$. Hence, 
eqs.\ (\ref{rhor.3}) and (\ref{conv}) yield the final result:
\begin{equation}
V_{\rm eff}(R)  =  N^2 \left(\frac{3}{4\pi R_g^2}\right)^{3/2} v_0 k_BT
           \exp\left(-\frac{3R^2}{4R_g^2}\right).
\label{final}
\end{equation}
The effective potential depends only on the magnitude $R = |{\bf R}|$ of the
center-to-center separation, because of rotational symmetry. 
Note that the interaction potential 
of eq.\ (\ref{final}) above, is identical to the Flory-Krigbaum 
effective interaction potential between the centers of mass of two
polymer chains \cite{flory:krigbaum:50}.  The fact that we are dealing
with dendrimers enters into the relation between $N$ and $R_g$. For long,
self-avoiding chains, $R_g \propto N^{3/5}$ holds; in the present study this is
not the case and we determine both quantities from experiment.

The important feature is that the resulting interaction has a Gaussian
form:
\begin{equation}
V_{\rm eff}(R) = \varepsilon\exp\left(-\frac{R^2}{\sigma^2}\right),
\end{equation}
with
\begin{equation}
\varepsilon = N^2 \left(\frac{3}{4\pi R_g^2}\right)^{3/2} v_0 k_BT
\label{epsilon}
\end{equation}
and
\begin{equation}
\sigma = \sqrt{\frac{4}{3}}R_g.
\label{sigma}
\end{equation}
The task is now to calculate $\varepsilon$ from the experimental data.
The number of monomers per dendrimer, 
is $N = 94$, (see Fig.\ \ref{dendrimer})
where the aromatic end-groups are also included in the counting as 
``single monomers''.

Next we need an estimate for the excluded volume parameter $v_0$. For this,
we set $v_0 = l_0^3$, where $l_0$ is the monomer length. Taking the 
realistic value $l_0 = 0.4\;{\rm nm}$ for the latter 
quantity, we obtain:
\begin{equation}
v_0 = 0.064{\;\rm nm^3}.
\end{equation}
The above values, together with the experimentally determined gyration
radius $R_g = 1.85\;$ nm yield:
\begin{equation}
\varepsilon = 10.42k_BT.
\label{numerical}
\end{equation}
The corresponding value for
{\it linear chains} is about $2k_BT$ \cite{ard1:00}.
Hence, the energy barrier for
the dendrimers at hand is about five times higher as that for
linear chains, a result that is physically reasonable as dendrimers
are more compact objects than chains. 
A general study of the structural and thermodynamic properties of
a system of particles interacting by a Gaussian potential
(the ``Gaussian core model'', GCM), has been presented
recently \cite{lang:etal:gcm}. 
The GCM shows
{\it no freezing} for $\varepsilon \lesssim 100$ \cite{lang:etal:gcm}, hence
we conclude that 
the system of dendrimers at hand will remain fluid at all 
concentrations.

Next, we express the density of dendrimers in units that are more
convenient.
Since there are $N_{\rm d}$
dendrimers in the volume $V$, the density $\rho_{\rm d}$ of dendrimers
is $N_{\rm d}/V$. 
In Ref.\ \cite{ballauff:99}, 
the volume fraction $\phi$ was used which is related to 
$\rho_{\rm d}$ by
\begin{equation}
\phi = \rho_{\rm d}V_p.
\label{phi}
\end{equation}
In order to suppress the dependence on $V_p$,
it is natural  
to use as the unit of length the parameter $\sigma$ of
the pair potential and to introduce a dimensionless measure of
the density, namely:
\begin{equation}
\eta = \frac{\pi}{6}\rho_{\rm d}\sigma^3,
\label{eta}
\end{equation}
which may be regarded as an effective volume 
fraction of the dissolved dendrimers.
From eqs.\ (\ref{sigma}), (\ref{phi}) and (\ref{eta}) we obtain:
\begin{eqnarray}
\eta & = &  
\frac{\pi}{6}\left(\frac{4}{3}\right)^{3/2}\frac{R_g^3}{V_p}\phi 
= 0.338\phi,
\label{etaphi}
\end{eqnarray}
where the known values $R_g = 1.85\;{\rm nm}$ and 
$V_p = 15.1\;{\rm nm}^3$ have been used \cite{ballauff:99}.

For every value of $\eta$, the Hypernetted
Chain (HNC) equation \cite{Hansen:86} for a system interacting via 
$\beta v(r) = 10.42\exp(-r^2/\sigma^2)$ was solved and the structure factor
$S(q)$ was obtained as a function of $q\sigma$. Using eq.\ (\ref{sigma})
we obtain $\sigma = 2.136\;{\rm nm}$. 
This is in close agreement with the effective radius 
of 2.4 nm estimated in Ref.\ \cite{ballauff:99} 
by a simple model valid only for dilute solutions. 
In what is to follow the $q$-scale was reexpressed
in ${\rm nm}^{-1}$ units. 

\section{Results and Discussion} 

\begin{figure}
\begin{center}
\includegraphics[width=10cm,height=10cm,angle=-90]
{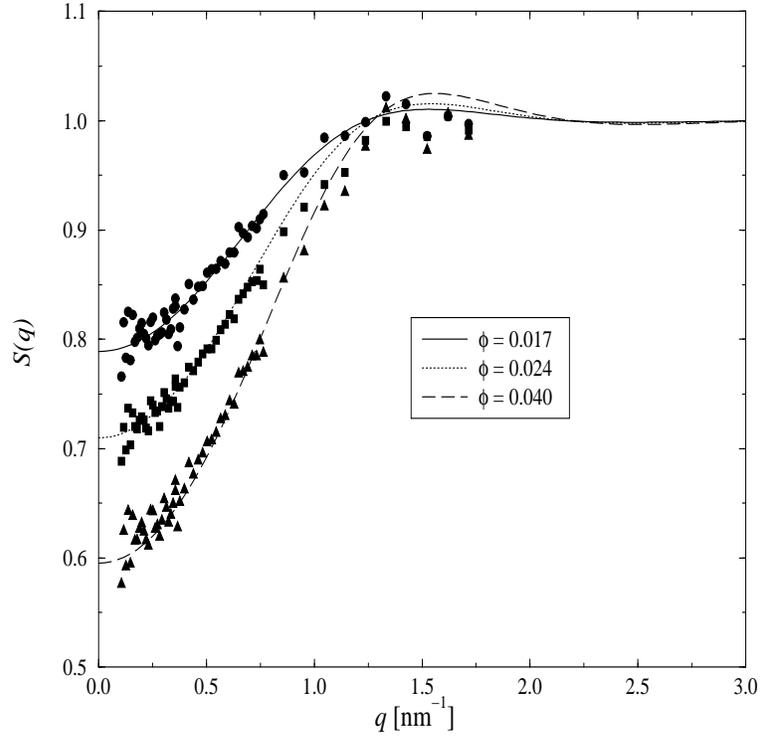}
\caption{Experimentally measured (points) and theoretically 
calculated (curves) structure factors $S(q)$ for solutions of 
dendrimers at three different concentrations. The theoretical curves
have been obtained by employing a Gaussian effective interaction 
between the dendrimers.}
\label{fits.gauss}
\end{center}
\end{figure}

As can be seen from Fig.\ \ref{fits.gauss}, the Gaussian interaction
potential yields excellent agreement between theory and experiment at
all three concentrations measured. Moreover, this is achieved without
the use of free fit parameters and it is physically reasonable
that soft, interpenetrable objects such as dendrimers interact by means 
of a correspondingly soft interaction. The boundedness of the effective
interaction at zero separations between the dendrimer centers is 
also physically correct: configurations where the centers of mass
of two dendrimers coincide 
are possible, without violation of 
excluded-volume conditions, hence the effective potential does not
diverge at the origin. This is quite analogous to the case
of polymer chains, where the effective interaction is {\it known}
to be Gaussian \cite{ard1:00}. 

The precise numerical value $\varepsilon = 10.42k_BT$
of the energy barrier, eq.\ (\ref{numerical}),
yields an optimal agreement with the experimental data.
However, small deviations from this
value can also be tolerated and by comparing with the SANS data we
have established the limits in which this parameter can vary as
$8.0 \lesssim \varepsilon \lesssim 12.0$, yielding a corresponding
tolerance interval $0.370\;{\rm nm} \lesssim l_0 \lesssim 0.425\;{\rm nm}$
for the ``effective monomer length'' $l_0$.

\begin{figure}
\begin{center}
\includegraphics[width=10cm,height=10cm,angle=-90]
{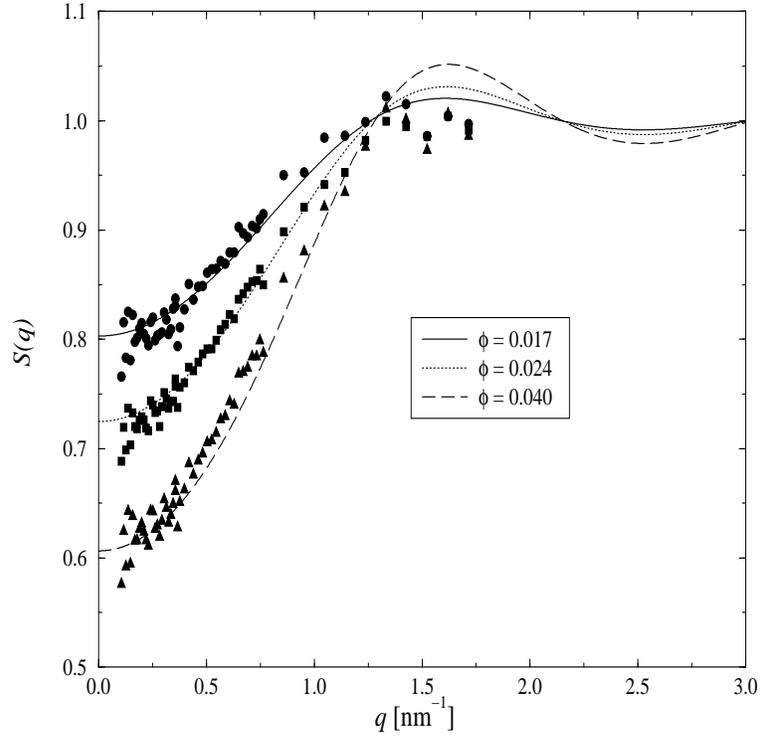}
\caption{Same as Fig.\ \ref{fits.gauss}, but now the theoretical
curves have been obtained by employing a hypothetical hard sphere (HS)
interaction
between the dendrimers, with a HS diameter $\sigma_{\rm HS} = 1.95R_g$.} 
\label{fits.hs}
\end{center}
\end{figure}

A prominent characteristic of the
Gaussian potential at the reduced temperatures considered here (and
also at higher ones), is its property to yield structure factors
which do not show any pronounced peak with increasing density, in
direct contradistinction with hard, diverging interactions such
as the hard sphere (HS) potential \cite{lang:etal:gcm}. In order 
to demonstrate this point, we have also attempted to fit the
experimental data using a hypothetical HS interaction between the
dendrimers, with the HS diameter $\sigma_{\rm HS}$ as a fit parameter.
The results are shown in Fig.\ \ref{fits.hs}. The best fit
at the lowest concentration is obtained by the choice
$\sigma_{\rm HS} = 1.95R_g$, which is thereafter kept 
constant. It can be seen that the fit quality worsens with increasing
density, as the HS interaction yields a too high peak as well as
a structure factor at low $q$-values which lies below the experimental
data. However, as we are in the very dilute regime, the strong
differences between the structure factors produced by these two
interactions (Gaussian and HS) are not very pronounced. 
These differences become
evident only if one looks at higher densities. 
The result found here is therefore in agreement with the 
findings of Ref.\ \cite{ballauff:99}. 
There it has been found that modeling the structure 
factor in terms of a simple hard-sphere 
ansatz leads to a satisfactory description of the 
data in the region of lowest volume fractions.

It is therefore evident that a more 
stringent test of theory can only be achieved  by 
SANS-studies conducted at much higher concentrations. 
Experiments using the G4 and G5 dendrimer 
used in Ref.\ \cite{ballauff:99,Potschke:99} are under way. Here
we present further evidence for the validity of the proposed interaction
by comparing with and by offering a theoretical explanation for
data already existing in the literature \cite{Ram:98,Amis:99}.

\begin{figure}
\begin{center}
\includegraphics[width=10cm,height=10cm,angle=-90]
{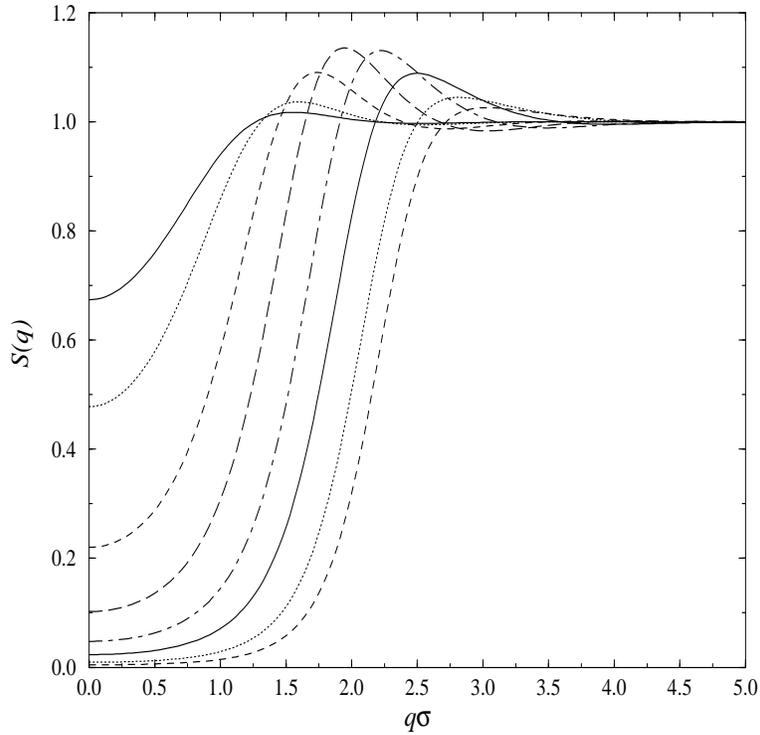}
\caption{The structure factor of the Gaussian core model at energy
$\varepsilon = 10k_BT$ as a function of the ``packing fraction''
$\eta$ given in eq.\ (\ref{eta}). From left to right:
$\eta = 0.01$, 0.02, 0.05, 0.10, 0.20, 0.40, 1.00 and 2.00.}
\label{gauss.anom}
\end{center}
\end{figure}

In Fig.\ 5 of Ref.\ \cite{Ram:98}, Ramzi {\it et al.}\ display the 
experimentally determined structure factor for a very wide concentration
range of dendrimers, ranging from dilute to above the overlap 
concentration. It is seen that $S(q)$ is deprived of any significant
structure; it has a weak peak of height $\sim 1.1$ at intermediate 
concentrations and thereafter the peak disappears and $S(q)$ becomes
a monotonic function of $q$.
At the same time,
this phenomenon is accompanied by a monotonic reduction of the 
osmotic compressibility
of the solution, the latter being proportional to the $q \to 0$ limit
of $S(q)$. This is {\it precisely} the behavior of $S(q)$ of the 
Gaussian core model \cite{lang:etal:gcm}. The latter has a 
freezing and reentrant melting transition 
with increasing density at energies $\varepsilon \geq 100k_BT$,
meaning that the liquid $S(q)$ has an anomalous dependence on the
density: the height of its principal maximum grows up to a certain
density and then it diminishes again. As the temperature is increased,
the anomaly in $S(q)$ remains, but the height of the principal peak 
becomes smaller. In order to provide a semi-quantitative comparison
with the data of Ref.\ \cite{Ram:98}, we show in Fig.\ \ref{gauss.anom}
the evolution of $S(q)$ of the GCM with density for an energy barrier
$\varepsilon = 10k_BT$ and 
packing fractions $\eta = 0.01 - 2.00$, corresponding to
$\phi \lesssim 0.67$. 
A striking similarity with the results of Ref.\ \cite{Ram:98} can 
be easily seen.

Similar conclusions hold when we compare our predictions with the
scattering data of Topp {\it et al.}, Figs.\ 7 and 8 
in Ref.\ \cite{Amis:99}. There, the $S(q)$'s from samples of two
different dendrimers are displayed; the same anomalous dependence 
of $S(q)$ on the concentration is seen, with the additional 
feature that the $S(q)$'s from the dendrimers with the larger
generation number have higher peaks than their counterparts
from the smaller dendrimers at the same concentration. In our
language, a larger generation number, which implies a larger 
monomer number $N$, corresponds to a higher energy barrier
$\varepsilon$, see eq.\ (\ref{epsilon}).
This automatically causes $S(q)$ to develop 
stronger peaks for larger generation numbers.

Of particular interest is the shape of $S(q)$ for very high concentrations
which, as can be seen in Refs.\ \cite{Ram:98} and \cite{Amis:99} 
is a monotonic function of $q$. This behavior can be fully understood
in the framework of the Gaussian interaction potential. It has been
recently shown that bounded potentials in general, show at high
densities a particular 
`mean-field behavior' \cite{lang:etal:gcm,likos:00,ard2:00}. The
direct correlation function $c(r)$ in this limit becomes equal
to $-\beta v(r)$, where $v(r)$ is the interaction potential at hand
and $\beta = \left(k_BT\right)^{-1}$.
For the case of the Gaussian potential, this implies that the
structure factor at the high density limit takes the 
form \cite{lang:etal:gcm,ard2:00}:
\begin{equation}
S(q) = {{1}\over{1 + \pi^{3/2}\beta\varepsilon\rho\sigma^3
                 \exp\left[-\left( Q\sigma/2 \right)^2\right]}},
\label{sofq.analytic}
\end{equation}
or, more generally for any bounded potential $v(r)$ \cite{likos:00,ard2:00}:
\begin{equation}
S(q) = \frac{1}{1+\beta\rho{\tilde v}(q)},
\label{sofq.general}
\end{equation}
where ${\tilde v}(q)$ is the Fourier transform of $v(r)$. Eq.\ 
(\ref{sofq.analytic}) shows that, for the GCM,
$S(q)$ has
the form of a ``smoothed out'' step function. It is small at low
$q$'s and approaches unity at high $q$'s, 
as ${\tilde v}(q) \to 0$ there.
The same conclusion holds, actually, for
any bounded potential whose Fourier transform is a
monotonically decaying, nonnegative function of $q$ \cite{likos:00},
as can be seen from eq.\ (\ref{sofq.general}).

The anomalous behavior of the peak heights of the structure factor
would be impossible if the assumed interaction between dendrimers were
diverging at the origin as any power law. Indeed, power-law systems
are known to undergo a freezing transition, this implying that the
corresponding $S(q)$ develops stronger and stronger peaks with
increasing density until, ultimately, freezing occurs when the
principal peak reaches the quasi-universal Hansen-Verlet value
$2.85$ \cite{hansen:verlet:69}. Hence, effective dendrimer-dendrimer
interactions with such kinds of divergence at the origin can be
immediately ruled out. This argument, however, does not rule out
{\it all} diverging interactions. Indeed, the peak-height anomaly
has also been observed in the framework of a theoretical
treatment of star polymers \cite{anom}, whose effective interaction
is diverging logarithmically at the origin. This behavior,
which has also been experimentally seen \cite{olaf:phd:95},  
is again intimately related to the reentrant melting behavior
of these systems \cite{watzlawek:etal:prl:99,martin:phd:00}.
In fact, it has already been suggested \cite{Amis:99} that dendrimers may 
resemble star polymers below the critical arm number 
$f_{\rm c} = 34$ at which crystallization is marginally 
possible \cite{watzlawek:etal:prl:99,martin:phd:00,witten:86}.
However, at variance with the Gaussian interaction, the star-star
effective potential \cite{Lik:98} does {\it not} have a mean-field
high density limit and the structure factor of stars never becomes
a monotonic function of $q$, even at extremely high densities
\cite{anom}, a feature which appears for bounded interactions only.
In this respect, there is a difference between the effective interaction
between the central monomers of two dendrites or two stars (which diverges) 
and that
between their centers of mass (which does not).

On purely theoretical grounds, the underlying
assumptions leading to the derivation of the logarithmic interaction
between star polymers do not hold for dendrimers. Unlike stars,
dendrimers do not obey a power-law dependence of the density profile
around the center; they show no self-similarity, captured in the
Daoud-Cotton blob model of the stars \cite{Daoud:CottonL82:1};
and they are at least one order of magnitude smaller than star polymers,
with the implication that many of the notions of polymer physics,
based on long chains and universality cannot be automatically carried
over to dendrimers. 

We conclude this section with a remark on the accuracy of the 
pair potential approximation. Though many-body forces 
between the dendrimers are necessarily
present in the solution, there are good reasons to believe that
their effect can be neglected to a very good approximation. This belief
is based on the one hand on the corresponding findings on star 
polymers \cite{triplet:00}. 
By a direct measurement of the triplet forces, it was
established there that these have a very small effect on the total
force on a star center, and this only shows up at concentrations considerably
beyond the overlap value. On the other hand, a recent simulational work
by Louis {\it et al.}\ \cite{ard1:00}, reached
a similar conclusion for single chains. There, it was found that the
pair potential yielding the correct thermodynamics of the system
remains density-independent up to the overlap concentration and displays
only a very weak density dependence above the latter. This density 
dependence is, of course, just another way to formulate the many-body
effects. Once more, they were found to be of minor significance.
Barring any dramatic alterations in the conformations of single
dendrimers upon increasing density (such as collapsing), we expect the
Gaussian pair potential picture to capture the salient characteristics of
the behavior of this system at all concentrations.  

\section{Summary and conclusions}
By using the density profile of an isolated dendrimer and the 
experimentally determined characteristics of the macromolecules
as input, we derived an effective interaction between the dendrimers'
centers of mass
which is Gaussian in form. By direct comparison with experimental data
at dilute solutions, we show that this interaction provides an
excellent description of the measured structure factor. Moreover, 
it reproduces correctly the as of now unexplained trends and features
of experimental structure factors at higher densities. A direct 
comparison with SANS data at high concentrations is, evidently, of
crucial importance for further putting the proposed theory into
a strong test. Work along these lines is currently in progress. 

\acknowledgments
The authors gratefully acknowledge financial support by the 
Deutsche Forschungsgemeinschaft and by the Bundesministerium 
f{\"u}r Forschung und Technologie.


\end{document}